\documentstyle[tighten,preprint,eqsecnum,epsfig,aps]{revtex}

\def\endproof{\nobreak\kern5pt\nobreak\vrule height4pt width4pt depth0pt}

\begin{document}
\draft
\hyphenation{
mani-fold
mani-folds
geo-metry
geo-met-ric
}

\title{Conditional
probabilities 
in Ponzano-Regge minisuperspace}

\author{Roman  Petryk\footnote{e-mail: petryk@physics.ubc.ca}}
\address{Department of Physics and Astronomy\\ University of British Columbia\\
Vancouver, BC V6T 1Z1, Canada}

\author{Kristin Schleich\footnote{e-mail: schleich@noether.physics.ubc.ca}} 
\address{Department of Physics and Astronomy\\ University of British Columbia\\
Vancouver, BC V6T 1Z1, Canada}

\date{\today}

\maketitle

\begin{abstract}

We examine the Hartle-Hawking no-boundary initial state for the Ponzano-Regge formulation 
of gravity in three dimensions. We consider the behavior of conditional probabilities 
and expectation values for geometrical quantities in this initial state for a simple 
minisuperspace model consisting of a two-parameter set of anisotropic geometries on a 
$2$-sphere boundary. We find dependence on the cutoff used in the construction of 
Ponzano-Regge amplitudes for expectation values of edge lengths. However, these expectation 
values are cutoff independent when computed in certain, but not all, conditional probability 
distributions. Conditions that yield cutoff independent expectation values are those that 
constrain the boundary geometry to a finite range of edge lengths. We argue that such 
conditions have a correspondence to fixing a range of local time, as classically associated 
with the area of a surface for spatially closed cosmologies. Thus these results may  hint 
at how classical spacetime emerges from quantum amplitudes.

\end{abstract}


\narrowtext

\section{Introduction}

Analysis of quantum behavior is usually guided by insight into the classical dynamics 
of the system. However, this connection is not  straightforward in the context of
quantum gravity.  Classical general relativity is explicitly described in terms of a 
coordinate system; in particular, the evolution of geometry is given as a function of a 
time coordinate. However, the quantum mechanics of gravity result in a wavefunction in 
which there is no explicit dependence on time. Rather, the classical idea of the time 
evolution of a geometry is expected to emerge from the form of the wavefunction itself. 
But how and under what conditions? Furthermore, how can information characterizing our 
classical spacetime such as homogeneity and isotropy be extracted from such a 
wavefunction?

Obtaining answers to such questions rests not only on knowing the correct quantum theory 
and initial conditions for the early universe, but also on understanding how to extract 
from the resulting wavefunction quantities that correspond to the properties of classical 
spacetime \cite{Hartle:1997hw}. Indeed, this issue has been key in the study of quantum 
cosmology. Much exploratory work on the likelihood of isotropy and other observed properties
of the universe has been done in the context of both semiclassical general relativity 
and minisuperspace models. (cf. \cite{Coleman:1991rs} and references therein.) In addition, 
various authors \cite{Hartle:1992as,Gell-Mann:1993kh,Griffiths:1984rx,Omnes:1994} 
have proposed an interpretation of quantum mechanics applicable to closed systems without 
external observers based on the consistent histories approach to quantum mechanics. This 
approach has been applied to the study of simple quantum systems related to gravity (cf.
\cite{Hartle:1993yv,Hartle:1994sg,Hartle:1997dc,Halliwell:2000yc}) with interesting results. 

These  suggestive avenues of research have been carried out in the context of a theory of 
quantum dynamics based on classical general relativity. However, it is generally accepted
that this theory is not itself a correct high energy description of quantum gravity; rather
it is expected to be an effective theory  that corresponds to the correct theory of quantum 
gravity only in the low energy limit. Furthermore, computations with this effective theory
can only be carried out in simple minisuperspace models, in the semiclassical limit, or with 
other approximation techniques. Therefore it is  useful to find other contexts in which we 
can probe the issues of how to associate the properties of classical spacetimes with quantum 
states. The Ponzano-Regge theory of gravity may provide one such context.

In 1968, Ponzano and Regge \cite{PR} noted an equivalence of the $6j$-symbols for spin to 
the 3-dimensional Regge action for gravity \cite{Regge:1961px}. They used this equivalence 
to formulate the partition function for 3-dimensional gravity in terms of a product of 
$6j$-symbols. This provided a well-formulated, calculable theory of 3-dimensional quantum 
gravity based on the sum over histories approach to quantum mechanics. Turaev and Viro 
\cite{TV} showed that a generalization of this theory formulated in terms of quantum 
$6j$-symbols provided new 3-manifold invariants. Ooguri \cite{Ooguri:1992ni} demonstrated 
that the Ponzano-Regge partition function is equivalent to Witten's 2+1 formulation of 
gravity \cite{Witten:1988hc} on closed orientable manifolds.  Barrett and Crane 
\cite{Barrett:1997gd} pointed out that the Biedenhorn-Elliot identity for $6j$-symbols
yields a discrete version of the Wheeler-de~Witt equation, thereby giving further insight
into the relation of  Ponzano-Regge theory to 3-dimensional gravity.

Clearly, Ponzano-Regge theory is  a potentially useful testing ground for 
the issues involved in the emergence of classical spacetime and the prediction of
isotropy from the wavefunction of the universe. As the theory  provides a large number of
geometrical degrees of freedom,  it yields quantum amplitudes for boundary geometries 
without any assumption of symmetry. Furthermore, its discrete nature means that
it is particularly amenable to numerical analysis. However, in order to implement such
studies, one must still understand how to formulate and interpret physical quantities in
this theory.

As a modest step toward this goal,  this paper examines the formulation of conditional 
probabilities and expectation values of geometrical quantities in a minisuperspace model 
of Ponzano-Regge theory. In particular, these quantities are computed in the Hartle-Hawking 
initial state (cf. \cite{Hartle:1983ai}) for simple two-parameter anisotropic boundary
geometries defined on the surface of a single tetrahedron. These models are useful in that
the expectation values can be calculated exactly. We find that a key issue in analyzing and 
interpreting such expectation values is the role of the cutoff used in the formulation of 
quantum amplitudes in Ponzano-Regge gravity.\footnote{The presence of a cutoff distinguishes 
Ponzano-Regge amplitudes from the Turaev-Viro formulation of 3-manifold invariants \cite{TV}
which  are by definition finite. } We find that the expectation values of edge lengths are 
linearly proportional to this cutoff. We also find  that certain, but not all,
expectation values of edge lengths that are conditioned on another edge length are
in fact cutoff independent. The factor determining  cutoff independence is whether or not
the condition imposed restricts the histories yielding the dominant contribution to the 
expectation value to those with support  on a finite region in the interior of the 
configuration space. We observe that such conditioned histories have a natural description 
in terms of intrinsic time. We conclude with a discussion of these results and their 
implications.

\section{Ponzano-Regge gravity}

We begin with the definitions of the $6j$-symbol and the Ponzano-Regge wavefunction.
This material is available widely in the literature (see, for example \cite{PR}, and the
generalization in \cite{TV}), but in varied notation; we summarize it here for convenience.

Let $j_{1}$, $j_{2}$, $j_{3}$, $j_{4}$, $j_{5}$, $j_{6}$ be non-negative integers or 
half-integers. An unordered $3$-tuple $(j_{a},j_{b},j_{c})$ is said to be {\it{admissible}} 
if the triangle inequalities, i.e., ${|j_{b}-j_{c}|}\le{j_{a}}\le{j_{b}+j_{c}}$ etc., are met 
and if the sum $j_{a}+j_{b}+j_{c}$ is an integer. The ordered $6$-tuple 
$(j_{1},j_{2},j_{3},j_{4},j_{5},j_{6})$ is said to be {\it{admissible}} if and only if each 
of the unordered $3$-tuples $(j_{1},j_{2},j_{3})$, $ (j_{3},j_{4},j_{5})$, 
$ (j_{5},j_{6},j_{1})$, $ (j_{2},j_{4},j_{6})$ are admissible. We can associate a $6j$-symbol
of $SU(2)$ with the admissible ordered $6$-tuple (cf. \cite{PR},  \cite{Racah}):
\begin{eqnarray}
\label{6j-equation}
\left\lbrace\matrix{j_{1}&j_{2}&j_{3}\cr j_{4}&j_{5}&j_{6}\cr}\right\rbrace={\sum_{z}}{\frac{(-1)^{z}(z+1)![{\Delta}(j_{1},j_{2},j_{3}){\Delta}(j_{3},j_{4},j_{5}){\Delta}(j_{5},j_{6},j_{1}){\Delta}(j_{2},j_{4},j_{6})]^{1\over{2}}}{(z-n_{1})!(z-n_{2})!(z-n_{3})!(z-n_{4})!(n_{5}-z)!(n_{6}-z)!(n_{7}-z)!}}
\end{eqnarray}
where
\begin{eqnarray}
n_{1}=j_{1}&+&j_{2}+j_{3} \ \ \ \ \ \  n_{2}=j_{3}+j_{4}+j_{5} \ \ \ \ \  \ n_{3}=j_{5}+j_{6}+j_{1}\nonumber\\
n_{4}&=&j_{2}+j_{4}+j_{6}\ \ \ \ \ \  \ \ \  \ \ n_{5}=j_{1}+j_{2}+j_{4}+j_{5}\nonumber\\
n_{6}&=&j_{2}+j_{3}+j_{5}+j_{6} \ \ \ \ \  n_{7}=j_{1}+j_{3}+j_{4}+j_{6}
\label{integer sums}
\end{eqnarray}
and the sum is over all non-negative integer values of $z$ resulting in non-negative arguments 
of the factorial. The function ${\Delta}(j_{i}, j_{j}, j_{k})$ is given by
\begin{equation}
\Delta(j_{i},j_{j},j_{k})={(j_{i}+j_{j}-j_{k})!(j_{i}+j_{k}-j_{j})!(j_{j}+j_{k}-j_{i})!\over{(j_{i}+j_{j}+j_{k}+1)!}} \, .
\end{equation}
For inadmissible 6-tuples, the $6j$-symbol is defined to vanish.

There is a natural geometric representation of the $6j$-symbol in terms of a $3$-dimensional 
tetrahedron $t$ (see Figure {\ref{fig:1}}). In particular, we can associate the edges of $t$ 
with the $j_{a}$'s in the $6j$-symbol as follows. Choose a face on $t$, label the edges of this 
face by $j_{1}$, $j_{2}$ and $j_{3}$. Next label the edge which shares no vertices with $j_{1}$ 
by $j_{4}$, the edge which shares no vertices with $j_{2}$ by $j_{5}$ and that which shares no 
vertices with $j_{3}$ by $j_{6}$. There are clearly two distinct choices of edge for $j_{2}$ 
or $j_{3}$---they determine the handedness of our labeling. The handedness, however, has no 
effect on calculations since the corresponding $6j$-symbols are equivalent according to the
symmetry
\begin{eqnarray}
\left\lbrace\matrix{j_{1}&j_{2}&j_{3}\cr j_{4}&j_{5}&j_{6}\cr}\right\rbrace=\left\lbrace\matrix{j_{1}&j_{3}&j_{2}\cr j_{4}&j_{6}&j_{5}\cr}\right\rbrace\ \, . \nonumber
\end{eqnarray}

The edge length associated with each $j_a$ is given by $l_{a}={j_{a}+{1\over2}}$, 
$a=1,2,...,6.$ Note that the requirement that the 3-tuples be admissible guarantees that
the edges $l_{a}$, $l_{b}$, $l_{c}$ form a closed triangle of non-zero area. The triangle 
inequalities do not, however, guarantee the associated tetrahedron has real volume; it is 
possible to construct {\it{hyperflat}} tetrahedra, that is tetrahedra with negative volume 
squared $V^2$ where 
\begin{eqnarray}
V^{2}={1\over{2^{3}(3!)^{2}}}\left|\matrix{0&l_{4}^{\ 2}&l_{5}^{\ 2}
&l_{6}^{\ 2}&1\cr l_{4}^{\ 2}&0&l_{3}^{\ 2}&l_{2}^{\ 2}&1\cr 
l_{5}^{\ 2}&l_{3}^{\ 2}&0&l_{1}^{\ 2}&1\cr l_{6}^{\ 2}
&l_{2}^{\ 2}&l_{1}^{\ 2}&0&1\cr 1&1&1&1&0\cr}\right| \, .
\label{volume}
\end{eqnarray}
This occurs when the sum of angles between three edges forming a vertex is greater than
$2\pi$. Clearly, tetrahedra with non-negative $V^2$ can be embedded in a 3-dimensional 
Euclidean space, while the same can not be done with tetrahedra with negative $V^2$. 
However, the ${V}^{2}<0$ tetrahedra can be  embedded in a 3-dimensional Lorentzian space 
(cf. \cite{Barrett:1994db}). 

For large edge lengths, the $6j$-symbol corresponds to the gravitational action through 
the relation
\begin{eqnarray}
\left\lbrace\matrix{j_{1}&j_{2}&j_{3}\cr j_{4}&j_{5}&j_{6}\cr}\right\rbrace&\simeq& \frac 1{\sqrt{12\pi V}}
\cos(I +\frac \pi{4})\nonumber \\
I&=&\sum_k l_k\theta_k \label{semi}
\end{eqnarray}
where $\theta_k$ is the angle between the outward-directed normals to the two faces that
intersect at edge $l_k$. $I$ is the contribution of the tetrahedron to the 3-dimensional
Regge action\cite{Regge:1961px} in units where $8\pi G=1$.

Next, we give the Ponzano-Regge partition function for compact 3-manifolds, or equivalently,
the Hartle-Hawking no-boundary initial state for Ponzano-Regge theory. Let $T[M]$ be a 
tessellation of a compact $3$-manifold $M$ with boundary ${\partial}M$. (A definition of a 
tessellation is provided in the Appendix.) Also let $T[{\partial}M]$  be the restriction of the
tessellation of $M$ to ${\partial}M$. Denote the sets of vertices, edges, triangles and 
tetrahedra in $T[M]$ as $S_{0}$, $S_{1}$, $S_{2}$ and $S_{3}$ respectively. Similarly 
denote the sets of vertices, edges,  and triangles in $T[{\partial}M]$ as $B_{0}$, $B_{1}$, 
$B_{2}$  respectively.  Let let $s_{m}$ be the cardinality of the set $S_{m}$ and  $b_{m}$
that of the set $B_{m}$. 

Next, let  $K$, the {\it{cutoff}},  be a non-negative integer or half-integer. Let  $\phi$  
be an assignment of a non-negative integer or half-integer value $j_{i}\le{K}$ to each edge 
in $S_{1}$. If every 6-tuple associated to a tetrahedron in $S_{3}$ by the assignment $\phi$ 
is admissible, then $\phi$ is termed an {\it{admissible assignment}}. Let $\{J_i \}$ denote 
the set of integer and half-integer values of edges in $B_1$. Note that $\{J_i \}$ is a 
subset of the set of all edges $\{j_i \}$.  Labeling the edges in the boundary
by indices $ 1, \ldots b_1$, the Ponzano-Regge wavefunction for a compact $3$-manifold 
is given by\footnote{Since non-admissible 6-tuples yield vanishing $6j$-symbols,
we could just as well sum over all 6-tuples instead of restricting ourselves to only 
admissible 6-tuples. } 
\begin{eqnarray}
\Psi[M,\{J_{i}\}]&=&{\lim_{{K}\to{\infty}}}\Psi_K[M,\{J_{i}\}]\nonumber\\
\Psi_K[M,\{J_{i}\}]&=&\sum_{\phi}\Lambda^{-(s_{0}-b_{0})}\prod_{i=1}^{b_{1}}(-1)^{J_{i}}(2J_{i}+1)^{{1\over{2}}}\prod_{i=b_{1}+1}^{s_{1}}(-1)^{2j_{i}}(2j_{i}+1)
\prod_{n=1}^{s_{3}}[t_{n}] \, .
\label{bounded ponzano regge}
\end{eqnarray}  
where  the sum is over all admissible assignments $\phi$, $[t_{n}]$ is 
\begin{equation}[t_{n}]=(-1)^{-(j_{a}+j_{b}+j_{c}+j_{d}+j_{e}+j_{f})}
\left\lbrace\matrix{j_{a}&j_{b}&j_{c}\cr j_{d}&j_{e}&j_{f}\cr}\right\rbrace
\end{equation}
in terms of the $6j$-symbol for the $n^{th}$ tetrahedron, and 
\begin{equation}
\label{lambda}
\Lambda=\sum_{p=0,{1\over2},...,K}(2p+1)^{2} \, .
\end{equation}
The term $\Lambda^{-(s_{0}-b_0)}$ regulates divergences in this expression
that appear due to the presence of internal vertices \cite{PR}. As a given $6j$-symbol 
is proportional to the contribution to the path integral amplitude for its associated 
tetrahedron, the product of $6j$-symbols is equivalent to the path integral amplitude 
for a given simplicial geometry. Thus each admissible assignment of edge lengths 
corresponds to a history in the path integral. The sum over admissible assignments 
is then a sum over histories for 3-d gravity. Note that we can associate a {\it{classically
allowed}} history as one composed of $6j$-symbols corresponding to tetrahedra of non-negative 
$V^2$, and a {\it{classically forbidden}} history as that containing one or more
tetrahedra of negative $V^2$. Finally, observe that the wavefunction will vanish if
there is no admissible assignment $\phi$.

For closed 3-manifolds, $\{ J_i\}=\emptyset$, and equation (\ref{bounded ponzano regge}) 
reproduces the usual partition function for Ponzano-Regge theory (cf. $\cite{PR,TV}$). 
Indeed, the wavefunction (\ref{bounded ponzano regge}) satisfies a natural composition law:
given tessellations  of two  $3$-manifolds $M$ and $N$ with the same boundary, 
$\partial M = \partial N$,  such that each boundary has identical tessellation, 
$T[{\partial}M]=T[{\partial}N]$, then
\begin{eqnarray}
\Psi[M\cup N]&=&{\lim_{{K}\to{\infty}}}\Psi_K[M\cup N]\nonumber\\
\Psi_K[M\cup N]&=&{\sum_{\{\{J_{i}\}|{J_{i}}\le{K}\}}}
\Lambda^{-b_{0}}{\Psi}_{K}[M,\{J_{i}\}]{\Psi}_{K}[N,\{J_{i}\}] \, . \label{gfn}
\end{eqnarray}
The inclusion of  $\Lambda^{-b_{0}}$ in the measure   regulates divergences appearing 
from the fact that the boundary vertices are now internal. This composition law also
reproduces the usual partition function for closed 3-manifolds. 

The expectation value of an operator $O$ is naturally formulated as
\begin{equation}
<{\cal O}>=\frac 1{\Psi[M\cup M]}{\lim_{{K}\to{\infty}}}
{\sum_{\{\{J_{i}\}|{J_{i}}\le{K}\}}}\Lambda^{-b_{0}}
{\Psi}_{K}[M,\{J_{i}\}]{ O}{\Psi}_{K}[M,\{J_{i}\}] \, . \label{expval}
\end{equation}
This definition can be viewed as the calculation of $O$ from the generating functional 
in Ponzano-Regge theory for which the initial and final states are identical.

An equivalent form of this expectation value  for suitably convergent numerator and 
denominator in (\ref{expval}),  is given by
\begin{eqnarray}
<{\cal O}>&=&{\lim_{{K}\to{\infty}}}<O>_K\nonumber \\
<O>_K&=&\frac {1}{\Psi_K[M\cup M]}
{{\sum_{\{\{J_{i}\}|{J_{i}}\le{K}\}}}\Lambda^{-b_{0}}
{\Psi}_{K}[M,\{J_{i}\}]{ O}{\Psi}_{K}[M,\{J_{i}\}]}\label{expval2}
\end{eqnarray}
where $\Psi_K[M\cup M]$ is given by (\ref{gfn}). It is important to note that both the 
above expressions are purely formal; the numerator and denominator may not necessarily 
converge in either. However, it is reasonable to expect that physically meaningful 
expectation values are convergent. Therefore, examination of the properties of such 
expressions as $K\to\infty$ is the subject of this paper. For notational convenience,
as all expectation values computed in the subsequent sections will be {\it{cutoff 
expectation values}} of form (\ref{expval2}) unless otherwise noted, we will suppress the 
subscript $K$, e.g. $<O>_K\equiv<O>$.

The simplest example of  the Hartle-Hawking no-boundary initial state is given by the 
tessellation of a  $3$-ball $B^3$ with $2$-sphere boundary as a single tetrahedron (see 
Figure {\ref{fig:1}}). The geometry of its boundary is completely specified by six independent 
edges. The wavefunction for this case is just
\begin{eqnarray}
\Psi[B^3,\{J_i\}]&=&{\lim_{{K}\to{\infty}}}\Psi_K[B^3,\{J_i\}] \nonumber \\
\Psi_K[B^3,\{J_i\}]&=&
\biggl[\prod_{i=1}^{6}(2J_{i}+1)^{{1\over{2}}}\biggr]\left\lbrace\matrix{J_1&J_2&J_3\cr J_4&J_5&J_6\cr}\right\rbrace \, .
\label{wavefunctionm}
\end{eqnarray}
The {\it{cutoff wavefunction}} and the limiting wavefunction are manifestly equivalent
in this case. However, although the wavefunction is cutoff independent, expectation values of
geometrical quantities such as edge lengths will not necessarily be so.

\section{Isotropic minisuperspace}
\label{ch2}

We will begin by considering the simplest minisuperspace model, that of the isotropic
boundary. In this model, we restrict all edges to have equal value  $l_X=X+{1\over2}$.
Then,  
\begin{equation}
\Psi[B^3,\{X\}]=\Psi_K[B^3,\{X\}]=
(2X+1)^{3}\left\lbrace\matrix{X&X&X\cr X&X&X\cr}\right\rbrace \, .
\label{wavefunction1}
\end{equation}
It follows from the admissibility conditions that $\Psi[B^3,\{X\}]$ vanishes unless $X$ is 
integer.  Figure {\ref{fig:2} displays this oscillatory wavefunction; it is apparent that it 
contains no striking structural features.

Next, we examine expectation values in this minisuperspace model. We do so by first 
calculating the cutoff expectation values in this model, then analyzing their behavior 
as $K\to\infty$. The cutoff expectation value of the edge length is given by
\begin{equation}
<l_{X}>={\sum_{X=0}^{K}{(X+{1\over2}){|{\Psi_K[B^3,\{X\}]}|^{2}}}
\over{\sum_{X=0}^{K}{|{\Psi_K[B^3,\{X\}]}|^{2}}}} \, .
\label{Type1lx}
\end{equation}
The uncertainty in the edge length
$
{\Delta}l_{X}=\sqrt{<(l_{X}-<l_{X}>)^{2}>} 
\label{delta l_{x}}
$
can be calculated similarly. We evaluate these expressions exactly using computer 
assistance. Figure {\ref{fig:3}} displays the results of these calculations
for values of ${0}\le{K}\le{200}$. The results of linear least-squares fits to these data 
yield\footnote{The respective coefficients of determination for each are ${R^{2}}\simeq{0.999935}$
and ${R^{2}}\simeq{0.99993}$, indicating good linear correlations. Least-squares fits 
to $2^{\rm{nd}}$ and $3^{\rm{rd}}$ order polynomials reveal that not only are the higher 
order coefficients at least four orders of magnitude smaller than those of linear order, 
but that there is also no significant improvement in $R^{2}$---an indication that higher 
order polynomial fits are unsuitable. Furthermore, least-squares fitting to non-polynomial 
functions of the form ${p_{1}}(K+{p_{2}})^{p_{3}}$ and $({p_{1}}K^{p_{3}}+{p_{2}})$, where 
$p_{1}$, $p_{2}$, $p_{3}$ are parameters to be determined by the fit, yield large off-diagonal 
asymptotic correlation matrix elements, suggesting the chosen fit functions are likely  
inappropriate due to the presence of too many degrees of freedom. Even so, $p_{3}$ for the 
$<l_{X}>$ data is $(0.999 \pm 0.001)$, while that for $\Delta l_{X}$ is $(0.996 \pm 0.001)$.
Linear fit functions are evidently the best choice.} 
\begin{eqnarray}
<l_{X}>&=&({0.8003}\pm{0.0003})K+({0.55}\pm{0.04})\nonumber\\ 
{\Delta}l_{X}&=&({0.16356}\pm{0.00007})K+({0.081}\pm{0.008}) \, .
\label{linear 2}
\end{eqnarray}

Observe that the linear dependence on the cutoff appears in computations of expectation 
values for the edge length of an isotropic tetrahedron embedded in flat 3-dimensional 
Euclidean space. As the volume of  a tetrahedron with edge $l$ is $V={l^3\over{6\sqrt{ 2}}}$, 
the unnormalized probability for the edge to have length  between $l$ and $l + dl$ is
$ dV = {l^2\over{2\sqrt{ 2}}}dl $. If we calculate the expected edge length and its 
uncertainty using this distribution cut off at a maximum edge length $K$, 
\begin{equation}
{<l>}=\frac 1{V_K} \int\limits_{0}^{K} \frac{l^3dl}{2\sqrt{ 2}} 
={{3}\over{4}}K \ \ \ {\rm{where}} \ \ \ V_K=\int\limits_{0}^{K}
\frac{l^2dl}{2\sqrt{ 2}} \, .
\label{envelope2}
\end{equation}
Similarly,
\begin{equation}
{{\Delta}{l}}={\sqrt{{3}\over{80}}}K \, .
\label{envelope6}
\end{equation} 
Furthermore, it is easy to see that the expectation
value for the edge length of any isotropic 3-dimensional solid will exhibit such linearity.

\section{Anisotropic minisuperspace}
\label{ch3}

Clearly, the isotropic edge length expectation values are cutoff dependent. But do there exist any
expectation values which are not dependent on $K$? We address such questions in this section using 
the simplest model possible to do so, the Hartle-Hawking initial state with boundary consisting 
of an anisotropic tetrahedron in which the six edges can take two distinct lengths. There are 
five types of this geometry characterized by the $j$-values and symmetries of the $6j$-symbol. 
Figure \ref{fig:4} illustrates the five types. 

Due to the formulation of the wavefunction in terms of edge lengths, the most convenient 
conditional expectation value to compute is  the expectation value of one edge length for 
a fixed value of the other. We will do so below for the type A tetrahedron, then
summarize the results for the remaining four types.

The type A tetrahedron has an equilateral triangle as its base and three isosceles triangles 
meeting at its vertex. The wavefunction is
\begin{equation}
\label{wavefunction2}
\Psi[ B^3,\{X,Y\}_{\rm{A}}]=\Psi_K[ B^3,\{X,Y\}_{\rm{A}}]= (2X+1)^{3\over{2}}(2Y+1)^{3\over{2}}
\left\lbrace\matrix{X&X&X\cr Y&Y&Y\cr}\right\rbrace \, ,
\end{equation}
where $X=0,1,2,...,K$, and $Y=0,{1\over{2}},1,...,K$. Figure \ref{fig:5} shows 
$\Psi[ B^3,\{X,Y\}_{\rm{A}}]$ as a function of $X$ and $Y$. Notice that
$\Psi[ B^3,\{X,Y\}_{\rm{A}}]$ vanishes for $X>2Y$ as such configurations are inadmissible, 
a feature not found in the isotropic case. This vanishing only occurs for certain classically 
forbidden histories, that is, those with $V^{2}<0$. This feature is easily understood in terms 
of the geometry of the tetrahedron; given the length $l_Y$ of the three edges forming the
peak of the tetrahedron, the triangle inequalities fix a maximum length for the three edges 
forming its base. 
 
Now the  minisuperspace expression for a cutoff expectation value, for example the edge 
length $l_{X}$, is given by
\begin{equation}
<l_{X}>={{\sum_{X,Y=0}^{K}(X+{1\over{2}})|\Psi_K[B^3,\{X,Y\}_{\rm{A}}]|^2}
\over{\sum_{X,Y=0}^{K}|\Psi_K[B^3,\{X,Y\}_{\rm{A}}]|^2}} \, ,
\label{unconstrained-1}
\end{equation}
where $X=0, 1, 2,..., K$ and $Y=0, {{1}\over{2}}, 1,..., K$. Exact evaluations of such 
expressions using computer assistance show that, as expected, the cutoff expectation values 
of edge lengths are again linear in $K$; Figure \ref{fig:6} displays these results for the 
type A tetrahedron and Table \ref{table2} gives the results of least squares fits to the 
data.\footnote{As in the isotropic case, we find that nonlinear fits are inappropriate.}
The slope of the linear dependence in $K$ differs somewhat from  that of the isotropic 
calculation for both $X$ and $Y$ edges as well as from each other. This difference is 
presumably due to the anisotropy of the tetrahedron which is explicitly reflected in the 
nested summation over the $X$ and $Y$ edges.

Table \ref{table1} summarizes configurations for the five different two-parameter anisotropic 
tetrahedral types. The computed unconditioned expectation values and uncertainties for all 
the anisotropic tetrahedra are linear in cutoff $K$. Table \ref{table2} displays the result 
of linear fits to these data. Note that in carrying out the 
evaluation of cutoff expectation values of form similar to (\ref{unconstrained-1}),
summations over $X$ and $Y$ may or may not include half-integer 
values depending on whether or not the admissibility conditions allow these configurations. 
Again, Table \ref{table1} summarizes the pertinent details. All unconditioned expectation 
values and uncertainties are linear with coefficients which vary slightly from each other 
and the isotropic values. Also observe that the coefficients of type B are equal, as expected 
from the symmetry under exchange of labeling $X $ and $Y$.

As for the isotropic case, the linear dependence on the cutoff $K$ appears in computations 
for all the anisotropic tetrahedra embedded in flat 3-dimensional Euclidean space. The 
volume of an anisotropic tetrahedron with edges $l_{X}$ and $l_{Y}$ is calculated as 
$V = \int P(l_{X}, l_{Y})dl_{X}dl_{Y}$, where $P(l_{X}, l_{Y})$ is a homogeneous polynomial
of degree one that vanishes identically for classically inadmissible configurations. The 
unconditional expected edge length $<l_{X}>$ is therefore $$ {<l_{X}>} = \frac{1}{V_{K}} \int\limits_{l_{Y}=0}^{K} \int\limits_{l_{X}=0}^{K} l_{X}  P(l_{X},l_{Y})dl_{X}dl_{Y} \propto K \ \ \ {\rm{where}} \ \ \ V_{K} = \int\limits_{l_{Y}=0}^{K} \int\limits_{l_{X}=0}^{K} P(l_{X},l_{Y})dl_{X}dl_{Y} \propto K^{3} \, . $$ Similarly, $$ \Delta l_{X} \propto K \, . $$

We next consider conditional expectation values for the anisotropic tetrahedron. Fixing 
a value of $l_Y$, the cutoff conditional expectation value of $l_{X}$ is given by
\begin{equation}
{<l_{X}|_{l_{Y}}>}={\sum_{X=0}^{K}{(X+{1\over2}){|{\Psi_K[B^3,\{X,Y\}_{\rm{A}}]}|}^{2}}
\over{\sum_{X=0}^{K}{|{\Psi_K[B^3,\{X,Y\}_{\rm{A}}]}|}^{2}}} \, ,
\label{anisotropic 1}
\end{equation}
where $X=0, 1, 2, ..., K$ and $l_Y=Y+\frac 12$ is a fixed integer or half-integer value.

One sees that, unlike $<l_{X}>$,  ${<l_{X}|_{l_{Y}}>}$ is independent of $K$ for edge 
lengths $l_Y$ such that $Y< \frac K2$ (see Figure \ref{fig:7}(b)). When cutoff independent, 
$<l_{X}|_{l_{Y}}>$ clearly exhibits linear dependence on $l_{Y}$ (see Figure \ref{fig:7}(a)). 
For $Y\ge \frac K2$, ${<l_{X}|_{l_{Y}}>}$ is cutoff dependent and also oscillates away from 
its previous linear relationship to $l_{Y}$ (as seen in Figure \ref{fig:7}(a)). Similarly, 
${\Delta}{l_{X}|_{l_{Y}}}$ is linearly dependent on $l_Y$ while cutoff $K$ independent 
and exhibits more complicated behavior for $Y \ge\frac K2$ (again, see Figure \ref{fig:7}). Note 
that ${<l_{X}|_{l_{Y}}>}$ and ${\Delta}{l_{X}|_{l_{Y}}}$ show little deviation from 
linearity for the classically forbidden values of $l_{Y}$. That is, it appears the 
functions are nearly constant for ${(K+{{1}\over{2}})} >{\sqrt{3}{l_{Y}}}$ where values 
$l_X >{\sqrt{3}{l_{Y}}}$ enter the summation. Classically forbidden histories therefore 
contribute comparatively little to these expectation values.

A study of the conditional expectation values and uncertainties for the remaining tetrahedral 
types verifies the correspondence of a constraining condition and a $K$ independent conditional 
expectation value for the appropriate range of the constraining edge length. In Table \ref{table1}
we give the ranges that the area-constraining edge lengths enforce. Note that both $X$ and $Y$ 
edges of the type B and C tetrahedra result in area-constraining conditions. 
The results of exact evaluations are displayed in Figures \ref{fig:9} and \ref{fig:10}. 
Observe that, when cutoff $K$ independent, the resulting expectation values and uncertainties 
are linear functions of the constraining edge length. 
Table \ref{table3} contains the results of linear fits to cutoff independent regions of the 
calculated values. All linear regions of conditional expectation values and uncertainties have
coefficients which vary slightly between the tetrahedral types. Also observe that the 
coefficients of type B are again equal, as expected from the symmetry under exchange of 
labeling $X$ and $Y$.

\section{Discussion}
\label{ch4}

When we examine the results for unconditioned expectation values for the isotropic
and anisotropic cases, we find explicitly linear cutoff dependence. However, when 
we look at conditioned expectation values, we find two strikingly different 
behaviors---sometimes the results are cutoff independent and linearly varying with 
the length of the conditioning edge, while at other times the calculated quantities 
are dependent on the cutoff and not linear in the conditioning edge length. We now 
interpret these behaviors.

\subsection{Unconditional Probabilities}

We already observed that the linearity in $K$ for the unconditioned cutoff
expectation values of edge length is what one would expect from a classical
probability distribution. This correspondence with a classical distribution is
not surprising given the importance of classical solutions in the continuum
formulation of 2+1 gravity \cite{Witten:1988hc}. However, this observation does not
address the role of the cutoff $K$ and its physical significance.

There are two natural interpretations. First, note that no physical scale 
has been set by  Ponzano-Regge gravity; the wavefunction contains no parameters other than 
that of the edge value $X$. In particular, observe that in the context of the semiclassical
limit (\ref{semi}),  a rescaling of the edge lengths $l_k\to\lambda l_k$ in the classical 
Regge action (\ref{semi}) results in a rescaling, $I \to \lambda I$. This scaling $\lambda$ 
can be absorbed  by redefining the length scale set by $8\pi G$. A change of edge length can 
thus be thought of as a change of this length scale to another value. Therefore, even though 
the gravitational constant is a length scale in the problem, one cannot vary it independently. 
Consequently there is no physical parameter to set a length scale that is independent of the 
cutoff $K$; $K$ is the only accessible scale parameter in the cutoff amplitudes. A change in $K$ 
in the calculation of this expectation value can therefore be viewed as a rescaling of units. 
As the cutoff expectation value scales linearly with $K$, its value is not physical.

Alternately, these results can also be interpreted in terms of intrinsic time. It is well-known 
that the position of a closed spatial hypersurface of a given geometry in a spacetime with 
closed spatial topology corresponds to a local time variable. For example, consider the maximal 
extension of 2+1 de Sitter spacetime. It has topology $S^2\times R$ and  metric 
$ds^2 = -dt^2 + {\alpha}^2 {\cosh}^2({\alpha}^{-1} t)d\Omega^2$ where $d\Omega^2$ is the unit 
round metric on the $2$-sphere. A round $2$-sphere of radius $R$ corresponds to only two spatial 
slices of this spacetime---those at $t= \pm \alpha \cosh^{-1}({\alpha}^{-1} R)$. Thus the spatial geometry of the $2$-sphere, in 
this case its area, encodes information about its local time position. In quantum cosmology, 
this classical association corresponds to an intrinsic time variable formed from the geometry 
of this spatially closed hypersurface (cf. \cite{Hartle:1984ut}, \cite{Kuchar:1980ht}).

In the case at hand, such a local intrinsic time would be associated with the area of the 
tetrahedron. For the isotropic tetrahedron, the area is proportional to the square of 
the edge length. In the context of quantum cosmology, expectation values of the type 
calculated here can  be thought of as averages over time. The choice of cutoff $K$ then 
corresponds to fixing a final time. Clearly, as the wavefunction (\ref{wavefunction1}) 
is not localized in $X$, unless the operators computed are themselves localized, their 
expectation values will depend on the cutoff. Here, the linear dependence of expectation 
values on $K$ indicates that this value does not reflect a geometrical property of the 
surface, but rather its time dependence.

Note that these interpretations, while somewhat different, are in fact compatible. 
The second simply associates the concept of final time to the cutoff of the problem.
Although this case of the isotropic boundary tetrahedron is particularly simple, it clearly 
delineates the relation in Ponzano-Regge theory.

Of course, one can also interpret the linear relation to $K$ as itself indicative of 
certain characteristics of the system. The correspondence of (\ref{linear 2}) to 
(\ref{envelope2}), (\ref{envelope6}) point to a correspondence of the quantum dynamics 
of the cutoff wavefunction to a classical evolution describing flat space. In this viewpoint, 
the fact that the coefficients $K$ in (\ref{linear 2})  differ slightly from those of the 
classical distribution (\ref{envelope2}), (\ref{envelope6}) might be expected from the oscillatory 
nature of the wavefunction (as displayed in Figure \ref{fig:2}). Such a view of cutoff expectation values treats the cutoff 
formulation of Ponzano-Regge theory as  relevant in its own right. However, this cutoff 
theory cannot be assumed to be equivalent to Ponzano-Regge theory. In particular, the 
tessellation independence of Ponzano-Regge amplitudes, a property explicitly dependent on 
the $K\to\infty$ limit, will not be achieved in the cutoff theory. Consequently, the linear 
relations observed in the cutoff theory may or may not reflect properties of Ponzano-Regge 
theory itself.

\subsection{Conditional Probabilities}

We now examine conditioned expectation values, focusing on ${<l_{X}|_{l_{Y}}>}$ for
the type A tetrahedral boundary. 
Again, the behavior of our results is easily understood from the classical geometry 
of the tetrahedron. Since all configurations with $X>2Y$ are inadmissible, these 
configurations have vanishing $\Psi[ B^3,\{X,Y\}_{\rm{A}}]$ (see Figure \ref{fig:5}) 
and thus vanishing probability 
density. Therefore, if $Y$ is fixed to be less than $\frac K2$, the conditional probability
density, and therefore ${<l_{X}|_{l_{Y}}>}$, will be independent of $K$. If $Y\ge \frac K2$, 
then there will be non-vanishing probability density for values of $X=K$. Consequently, the 
expectation value will exhibit cutoff dependence if they include configurations in the range 
$Y\ge \frac K2$. Figure \ref{fig:7} displays the results of these calculations.
 
Interpreting in terms of intrinsic time, fixing $l_Y$ in the calculation of the expectation 
value restricts the  histories that contribute to (\ref{anisotropic 1}) to those 
corresponding to a particular range of edge lengths, or equivalently areas of the 
tetrahedron. As this area corresponds classically to intrinsic time,  this restriction 
amounts to selecting a fixed range of intrinsic time, then computing the time average of 
the expectation value over this range. Therefore, ${<l_{X}|_{l_{Y}}>}$ yields a result 
that reflects properties of an edge length associated with a particular bounded time 
average for a suitable range of $Y$. For this range, the limit $K\to\infty$ of 
${<l_{X}|_{l_{Y}}>}$ is well-defined. Therefore such conditional expectation values yield 
a physically meaningful result. 

Interestingly enough, ${<l_{X}|_{l_{Y}}>}$, when well-defined, is linear in $l_Y$ (as seen
in Figure \ref{fig:7}). Now one can repeat the arguments on the lack of another physical 
scale parameter in the problem to argue that a change in $l_Y$ corresponds to a change of 
scale. This again indicates that the value of ${<l_{X}|_{l_{Y}}>}$  is scale dependent. 
However, the situation is quite different from the case of the unconditioned calculations.
Now, one has independent control over the physical scale and the cutoff needed in the 
definition of the Ponzano-Regge theory. Therefore, the linearity seen in these expectation 
values  can be taken as physical. Furthermore, one can argue that  the linearity of 
${<l_{X}|_{l_{Y}}>}$ in $l_Y$ demonstrates a physical property of this amplitude, namely 
that that this quantum amplitude corresponds to a flat space. 

Similar to analysis of the unconditioned case, linear dependence on the conditioning 
edge $l_{Y}$ appears in computations for tetrahedra embedded in flat 3-dimensional 
Euclidean space. Since the volume of a type A tetrahedron with edges $l_{X}$ and $l_{Y}$ 
is $V = \frac{{\, l_{X}}^2}{12} \sqrt{3{{l_{Y}}^2}-{{l_{X}}^2}} $, the unnormalized 
probability for the edge to have length  between $l_{X}$ and $l_{X} + dl_{X}$ is
$$ dV = \left| \, \frac{\, l_{X}}{6} (3{{l_{Y}}^2}-{{l_{X}}^2})^{\frac{1}{2}} - \frac{{\, l_{X}}^3}{12} (3{{l_{Y}}^2}-{{l_{X}}^2})^{-\frac{1}{2}} \, \right| \, dl_{X} \, . $$ 
Calculating the conditional expected edge length $<l_{X}|_{l_{Y}}>$ and uncertainty 
$\Delta l_{X} |_{l_{Y}}$ we obtain 
\begin{equation}
{<l_{X}|_{l_{Y}}>} = \frac 1{V_{l_{Y}}} \int\limits_{0}^{\sqrt{3} l_{Y}} l_{X} dV \simeq 1.23 l_{Y} \ \ \ {\rm{where}} \ \ \ V_{l_{Y}} = \int\limits_{0}^{\sqrt{3} l_{Y}} dV \, ,
\label{envelope2_b}
\end{equation}
and similarly
\begin{equation}
{{\Delta}{l_{X}}|_{l_{Y}}} \simeq 0.498 l_{Y} \, . 
\label{envelope6_b}
\end{equation}
Again observe that the constants of proportionality for the classical calculation are close 
to those of the corresponding Ponzano-Regge calculation. As for the unconditioned case, the 
differences can be attributed to the oscillatory nature of the wavefunction 
$\Psi[ B^3,\{X,Y\}_{\rm{A}}]$ (as shown in Figure \ref{fig:5}).

If the key property yielding cutoff independence is that the condition fixes a bounded
range of intrinsic time, then it follows that $<l_{Y}|_{{l_{X}}}>$ and 
${\Delta}l_{Y}|_{{l_{X}}}$ should not exhibit this cutoff independence for any range of 
$l_X$. Geometrically, fixing $l_X$ does not restrict the range of $l_Y$. Hence, this 
condition will not appropriately restrict the region of non-vanishing probability 
density. Calculations of these quantities verify this hypothesis (see Figure \ref{fig:8}). 
The expectation value $<l_{Y}|_{{l_{X}}}>$ exhibits cutoff dependence for all values of 
$l_X$. The fact that this expectation value is $K$ dependent implies that it is not 
well-defined in the $K\to\infty$ limit. Hence the hypothesis that only conditional 
expectation values for which conditions restricting the range of fixed intrinsic time 
result in physically meaningful answers is verified. Furthermore, observe that the 
linear correlations displayed between the cutoff $K$ independent conditional calculations 
and the constraining edge length are not present in either $<l_{Y}|_{{l_{X}}}>$ or 
${\Delta}l_{Y}|_{{l_{X}}}$.\footnote{Least-squares fits on these data quantify this 
absence of linear correlation. The coefficients of determination for linear fits are
respectively $R^2 \simeq 0.1$ and $R^2 \simeq 0.4$.}

A study of the conditional expectation values and uncertainties for the other anisotropic 
tetrahedral types verifies this behavior---when cutoff $K$ independent, the resulting 
expectation values and uncertainties are linear functions of the constraining edge length. 
Again, Table \ref{table1} gives the ranges that the area-constraining edge lengths enforce. 
Recall that both $X$ and $Y$ edges of the type B and C tetrahedra result in area-constraining
conditions. Once more, Table \ref{table3} contains the resulting linear fits for the appropriate 
ranges of the constrained edge lengths for all five anisotropic types. These results are 
consistent with the flat space interpretation as for the type A tetrahedron.

\section{Conclusions}

This simple study of expectation values on Ponzano-Regge minisuperspace demonstrates
the utility of this theory in addressing issues of the emergence and properties of
classical spacetime. It is clearly not true that all expectation values that can be
formulated for a fixed cutoff value are physically meaningful. It is also clear that not
all conditions on such expectation values will yield a cutoff independent result. However,
what is significant is the close correspondence between the formulation of conditional
expectation values that are cutoff independent quantities---and thus physically
meaningful---and conditions that correspond to fixing a finite range of time. Furthermore,
the linear nature of these cutoff independent quantities on the constraining edge length
is consistent with embedding of the tetrahedron in the classical flat background. These
results are not just of mathematical interest, but indicate that Ponzano-Regge theory may
be able to provide insight into the formulation of physically interesting expectation
values.

Clearly, one can ask whether or not more can be determined from well-defined conditioned
expectation values. In particular, expectation values of the form ${<l_{X}|_{l_{Y}}>}$
loosely correspond to  the characterization of the isotropy of a universe at a given time.
The interesting fact about the results for ${<l_{X}|_{l_{Y}}>}$ is that, as seen in Table
\ref{table3}, they are dependent on the boundary tetrahedron type. Therefore, it is not
clear that the coefficients of $l_Y$ hold much meaning in and of themselves. However, if
one chooses to interpret the values relative to the isotropic tetrahedron, it is clear
that $\frac{l_Y}{2} < {<l_{X}|_{l_{Y}}>}< \frac{3l_Y}{2} $. Thus, although the anisotropy
of the surface in this interpretation is not zero, it is in some sense small.

What is interesting about these results is the direction in which they point. Ideally, one
would like to be able to search for quantities characterizing the emergence of classical
spacetime in situations for which it is not so clear what the relevant degrees of freedom
are. So what characterizes whether or not the expectation values computed are relevant?
From our results, we see that requiring cutoff independence in conditional expectation
values is crucial and required for physically meaningful results. Furthermore, these results
are compatible with the presence of a localization in intrinsic time.

Given Ooguri's work \cite{Ooguri:1992ni} demonstrating the equivalence of 2+1 gravity in the
Witten variables to Ponzano-Regge theory, it is clearly of interest to ask how our results
fit into the issue of appearance of time in 2+1 gravity. To this end, note that, in the
continuum theory, the phase space of $SO(2,1)$ holonomies parameterizes the inequivalent
flat spacetimes for different topologies in 2+1 gravity. The wavefunctions for 2+1 gravity
are functions of suitable projections of this phase space. That is the basis of Witten's
approach---he quantizes gravity using diffeomorphism invariant $SO(2,1)$ holonomy variables
\cite{Witten:1988hc}. His approach results in a vanishing Hamiltonian and gauge invariant
states. Witten's quantization is therefore time independent. Alternately, Moncrief
\cite{Moncrief:1989dx}, and Hosoya and Nakao \cite{Hosoya:1989yj} have given a quantization
based on Arnowitt-Deser-Misner variables and the York decomposition. In that formulation
they use a time variable defined as the trace of the extrinsic curvature $K$.\footnote{Note
that although $K$ is referred to as the {\em{extrinsic time}}, it is however a phase space
variable and is therefore intrinsic to the description.} The formulation yields a non-vanishing
Hamiltonian and states that are dependent on $K$, i.e. are time dependent. Now, in
\cite{Carlip:1990kp} Carlip compares the Witten quantization to that due to Moncrief, Hosoya
and Nakao. Carlip shows that these two quantizations are exactly equivalent for the case of the
two-torus, and explicitly constructs a canonical transformation between Moncrief's and Witten's
variables for that case. Carlip thus demonstrates the possibility of canonically transforming
an explicitly time dependent quantum theory of 2+1 gravity to one in which the dependence is
hidden in the new variables.

How do our results fit into this picture? Our analysis of Ponzano-Regge theory through the
calculation of expectation values is formulated with respect to a fixed boundary. Clearly
fixing a boundary corresponds to a fixed slicing. Thus our (discrete) formulation parallels
that of Moncrief, Hosoya and Nakao; we expect that our amplitudes to be time dependent (i.e.
cutoff dependent) and indeed they are. We found that unconditional expectation values are
linear in the cutoff, precisely the dependence that these values have in the classical
probability distribution for flat space. The linearity of expectation values is also
consistent with uniqueness of the solution for $2$-sphere topology in 2+1 gravity
\cite{Witten:1988hc}. As the phase space for the $2$-sphere consists of a point, the
wavefunction in this formulation is trivial.

However, we also demonstrate that we can extract from these amplitudes physical behavior
that is in fact time independent by considering certain conditional expectation values.
More convincingly, we found that well-defined conditional expectation values are linear
functions of the conditioning edge length. Again this behavior reflects the underlying
flat space structure we expect. Part of the utility of our method is that we can extract
physical behavior without forming maps between gauge independent and gauge dependent
quantizations.

We carried out this analysis in terms of simple minisuperspace models in which all calculations
could be carried out exactly. It is important to note that the computation of expectation
values in the anisotropic minisuperspace approximation involved a restriction of the sum over
edge lengths from six values to two. This restriction will result in a reduction of the degree
of divergence of such expressions if no further changes in the measure are carried out. This
feature is common to other minisuperspace models as freezing out degrees of freedom typically
results in the redefinition of the measure. Clearly, this point indicates that the results we
have found are likely qualitative, rather than quantitative. Even so, the issue of the
divergence of the expectation values with cutoff will remain in the full Ponzano-Regge theory.
It is clearly of interest to study how such  results will appear. Given what our minisuperspace
results indicate, one can imagine formulating similar conditional expectation values in the full
Ponzano-Regge theory, numerically evaluating these using Monte Carlo techniques, and searching
for cutoff independence as a signal of a meaningful expectation value with a physical
interpretation. Clearly, it is of interest to see whether such conditional measurements
have any connection to the emergence of classical spacetime.

Furthermore, it is clearly important to consider such conditional expectation values in the
context of other 3-manifold topologies. Ionicioiu and Williams \cite{Ionicioiu:1998ti}
computed amplitudes for 3-manifolds in a closely related theory of Turaev and Viro \cite{TV}
in which 3-manifold invariants are formulated in terms of a topological field theory. Ooguri
\cite{Ooguri:1992ni} related the formulation of the Hartle-Hawking no-boundary initial state
on handlebodies to that of a Chern-Simons theory.  It is obviously of interest to study the
characterization of such solutions in the context of the interpretation of quantum cosmology
as their solution space is no longer zero dimensional. So whether or not the Hartle-Hawking
wavefunction yields conditional expectation values  with special properties is clearly of
interest.

\acknowledgements

We would like to thank Gordon Semenoff for useful conversations. This work was 
partially supported by the Natural Sciences and Engineering Research Council of 
Canada. R.~P. also acknowledges Canada Foundation for Innovation funding of
the Beowulf cluster \rm{vn.physics.ubc.ca}, used to compute the presented results.

\appendix

\section{Definitions}

In order to concretely define a tessellation, we begin by defining simplices
\cite{spanier}: Let the {\it{vertices}} $v_1,v_2,\ldots ,v_{n+1}$ be affinely
independent points in ${\bf R}^{m}$ where $m \ge n+1$. An {\it n-simplex}
$\sigma^n$ is the convex hull of these points:
$$ \sigma^n = \{x \in {\bf{R}}^m|
x=\sum_{i=1}^{n+1}\lambda_i v_i; \, \lambda_i \ge 0;\ \sum_{i=1}^{n+1}
\lambda_i =
1 \} \, .$$
A 0-simplex is a point or vertex, a 1-simplex
is a line segment or edge, a 2-simplex is a triangle including its interior and
a 3-simplex is a solid tetrahedron. A simplex constructed from a subset of the
vertices is called a {\it face}. For example, the 0-simplices or vertices of an
n-simplex are all faces of that simplex. Similarly, 1-simplices or edges formed
from any two distinct vertices of an n-simplex are also faces of that simplex.
 
Next note that a {\it simplicial complex} $S$ is a topological space $|S|$ and
a set of simplices $S$ such that
\hfill\break
$\ \ $i) $|S|$ is a closed subset of some finite dimensional Euclidean space.
\hfill\break
$\ \ $ ii) If $F$ is a face of a simplex in $S$, then $F$ is also contained in
$S$.
\hfill\break
$\ \ $ iii) If $B,C$ are simplices in $S$, then $B \cap C$ is either empty or a
face of both $B$ and $C$. \hfill\break
\noindent The topological space $|S|$ is the union of all simplices in
$S$.\par
 
Recall that a homeomorphism is a continuous, invertible map between topological
spaces. Then
 
\proclaim Definition. A {\it tessellation} $T[M]$ is a homeomorphism $T:M \to
|S|$ from an
n-manifold $M$ to the topological space given by the simplicial
complex $S$.
 
In other words, homeomorphisms $T$  tessellate n-manifolds into a collection of
n-simplices. A tessellation of $M$ is not unique---different maps $T$ can map
the same manifold $M$ to different  simplicial complexes. That is, we can choose an
appropriate $T$ to map the manifold to as few or as many n-simplices as we
desire.

\begin{table}[htbp]
\begin{center}
\begin{tabular}{ c c c c c c c c }
type & $J_{1}$ & $J_{2}$ & $J_{3}$ & $J_{4}$ & $J_{5}$ & $J_{6}$ &
restrictions \\ \hline
A & X & X & X & Y & Y & Y &
$ X \le 2Y$ \ ,
\ \ \ X \ {\rm{integer}} \\ \hline
B & X & X & Y & X & Y & Y &
$X \le 2Y \ ,
\ \ \ Y \le 2X$ \ ,
\ \ \ X \ {\rm{integer}} \ ,
\ \ \ Y \ {\rm{integer}} \\ \hline
C & X & X & Y & Y & Y & Y &
$X \le 2Y \ ,
\ \ \ Y \le 2X$ \ ,
\ \ \ X \ {\rm{integer}} \ ,
\ \ \ Y \ {\rm{integer}} \\ \hline
D & X & Y & Y & X & Y & Y &
$X \le 2Y$ \ ,
\ \ \ X \ {\rm{integer}} \\ \hline
E & X & Y & Y & Y & Y & Y &
$X \le 2Y$ \ ,
\ \ \ X \ {\rm{integer}} \ ,
\ \ \ Y \ {\rm{integer}} \\
\end{tabular}
\end{center}
\caption{The five distinct two-parameter anisotropic types and
additional restrictions on edges due to the triangular
inequalities and integer valued sum of each 3-tuple.}
\label{table1}
\end{table}

\begin{table}[htbp]
\begin{center}
\begin{tabular}{ c c c c }
type & $ l_{i} $ & $< l_{i} > $ & $ \Delta l_{i}$ \\ \hline
A & 
$ l_{X} $ & 
$ ({0.6871}\pm{0.0002})K+({0.52}\pm{0.02}) $ & 
$ ({0.23031}\pm{0.00005})K+({0.137}\pm{0.006}) $ \\ 
 & 
$ l_{Y} $ & 
$ ({0.75069}\pm{0.00003})K+({0.554}\pm{0.004}) $ & 
$ ({0.17435}\pm{0.00002})K+({0.127}\pm{0.002}) $ \\ \hline
B & 
$ l_{X} $ & 
$ ({0.7474}\pm{0.0002})K+({0.53}\pm{0.02}) $ & 
$ ({0.16991}\pm{0.00004})K+({0.106}\pm{0.005}) $ \\ 
 & 
$ l_{Y} $ & 
$ ({0.7474}\pm{0.0002})K+({0.53}\pm{0.02}) $ & 
$ ({0.16991}\pm{0.00004})K+({0.106}\pm{0.005}) $ \\ \hline 
C &
$ l_{X} $ & 
$ ({0.7072}\pm{0.0002})K+({0.52}\pm{0.02}) $ & 
$ ({0.18966}\pm{0.00005})K+({0.126}\pm{0.005}) $ \\ 
 &
$ l_{Y} $ & 
$ ({0.7516}\pm{0.0002})K+({0.55}\pm{0.02}) $ & 
$ ({0.18001}\pm{0.00004})K+({0.119}\pm{0.005}) $ \\ \hline
D &
$ l_{X} $ & 
$ ({0.5468}\pm{0.0001})K+({0.73}\pm{0.02}) $ & 
$ ({0.28722}\pm{0.00007})K-({0.047}\pm{0.008}) $ \\ 
 &
$ l_{Y} $ & 
$ ({0.79883}\pm{0.00002})K+({0.582}\pm{0.002}) $ & 
$ ({0.15267}\pm{0.00001})K+({0.108}\pm{0.002}) $ \\ \hline
E & 
$ l_{X} $ & 
$ ({0.5310}\pm{0.0001})K+({0.36}\pm{0.01}) $ & 
$ ({0.29080}\pm{0.00007})K+({0.236}\pm{0.008}) $ \\ 
 & 
$ l_{Y} $ & 
$ ({0.7926}\pm{0.0002})K+({0.57}\pm{0.02}) $ & 
$ ({0.16197}\pm{0.00004})K+({0.102}\pm{0.004}) $ 
\end{tabular}
\end{center}
\caption{Unconditional cutoff expectation values and uncertainties for the
two-parameter anisotropic types.}
\label{table2}
\end{table}

\begin{table}[htbp]
\begin{center}
\begin{tabular}{ c c c c c }
type & $ l_{i} $ & conditioning & \hspace{-0.7cm} $ < l_{i} |_{l_{c}} > $ & \hspace{-0.0cm} $\Delta l_{i} |_{l_{c}} $ \vspace{-0.1cm} \\ 
 & & edge $l_{c}$ & & \\ \hline
A & 
$ l_{X} $ & 
$ l_{Y} $ & 
\hspace{-0.7cm} $ ({1.3601}\pm{0.0005})l_{Y}+({0.04}\pm{0.03}) $ & 
\hspace{-0.0cm} $ ({0.387}\pm{0.002})l_{Y}+({0.0}\pm{0.1}) $ \\ \hline
B & 
$ l_{X} $ & 
$ l_{Y} $ & 
\hspace{-0.7cm} $ ({1.316}\pm{0.003})l_{Y}+({0.0}\pm{0.2}) $ & 
\hspace{-0.0cm} $ ({0.298}\pm{0.002})l_{Y}-({0.04}\pm{0.09}) $ \\ 
& 
$ l_{Y} $ & 
$ l_{X} $ & 
\hspace{-0.7cm} $ ({1.316}\pm{0.003})l_{X}+({0.0}\pm{0.2}) $ & 
\hspace{-0.0cm} $ ({0.298}\pm{0.002})l_{X}-({0.04}\pm{0.09}) $ \\ \hline 
C &
$ l_{X} $ & 
$ l_{Y} $ & 
\hspace{-0.7cm} $ ({1.5046}\pm{0.0001})l_{Y}+({0.010}\pm{0.006}) $ & 
\hspace{-0.0cm} $ ({0.4202}\pm{0.0001})l_{Y}-({0.011}\pm{0.007}) $ \\ 
&
$ l_{Y} $ & 
$ l_{X} $ & 
\hspace{-0.7cm} $ ({1.6224}\pm{0.0009})l_{X}-({0.03}\pm{0.06}) $ & 
\hspace{-0.0cm} $ ({0.346}\pm{0.001})l_{X}-({0.04}\pm{0.08}) $ \\ \hline
D &
$ l_{X} $ & 
$ l_{Y} $ & 
\hspace{-0.7cm} $ ({0.906}\pm{0.002})l_{Y}+({0.2}\pm{0.1}) $ & 
\hspace{-0.0cm} $ ({0.430}\pm{0.001})l_{Y}-({0.20}\pm{0.06}) $ \\ \hline
E & 
$ l_{X} $ & 
$ l_{Y} $ & 
\hspace{-0.7cm} $ ({1.10294}\pm{0.00003})l_{Y}-({0.0250}\pm{0.002}) $ & 
\hspace{-0.0cm} $ ({0.5329}\pm{0.0001})l_{Y}+({0.022}\pm{0.006}) $ 
\end{tabular}
\end{center}
\caption{Conditional expectation values and uncertainties for the two-parameter
anisotropic types.}
\label{table3}
\end{table}

\begin{figure}[htbp]
\epsfxsize=2.5in
\centerline{\epsfbox[10 0 600 760]{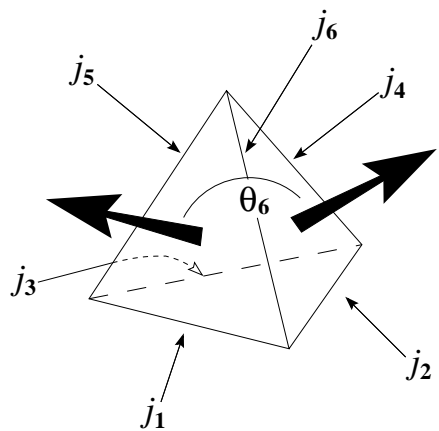}}
\caption{A tetrahedral representation of the $6j$-symbol.}
\label{fig:1}
\end{figure}

\begin{figure}[htbp]
\epsfxsize=3.0in
\centerline{\epsfbox[10 100 600 860]{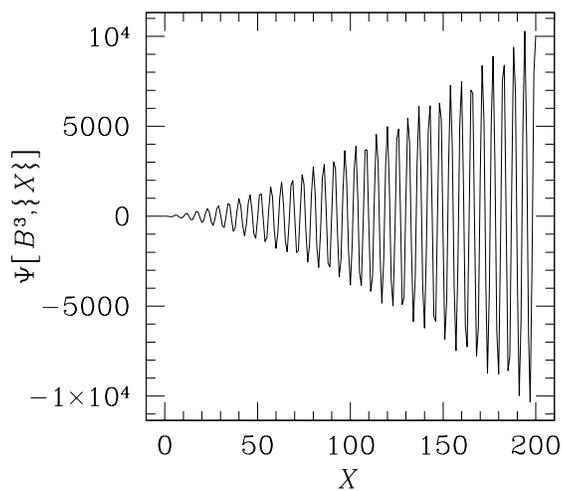}}
\caption{The isotropic wavefunction $\Psi[B^3,\{ X \}]$ for the range
${0}\le{X}\le{200}$.}
\label{fig:2}
\end{figure}

\begin{figure}[htbp]
\epsfxsize=3.0in
\centerline{\epsfbox[10 100 600 1260]{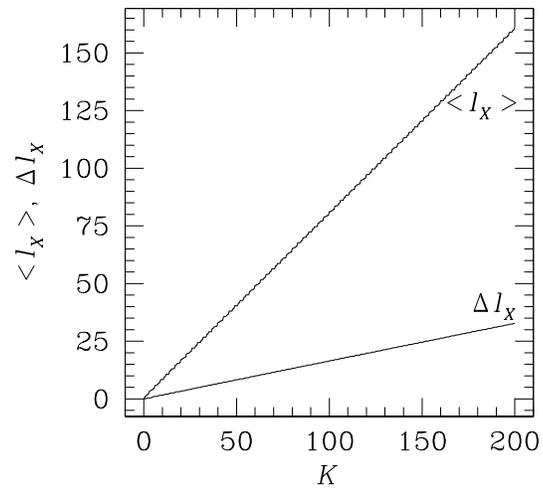}}
\caption{The cutoff expectation value $<l_{X}>$ and uncertainty
${\Delta}l_{X}$ for the isotropic boundary for the range
${0}\le{K}\le{200}$.}
\label{fig:3}
\end{figure}

\begin{figure}[htbp]
\epsfxsize=4.75in
\centerline{\epsfbox[10 -50 600 910]{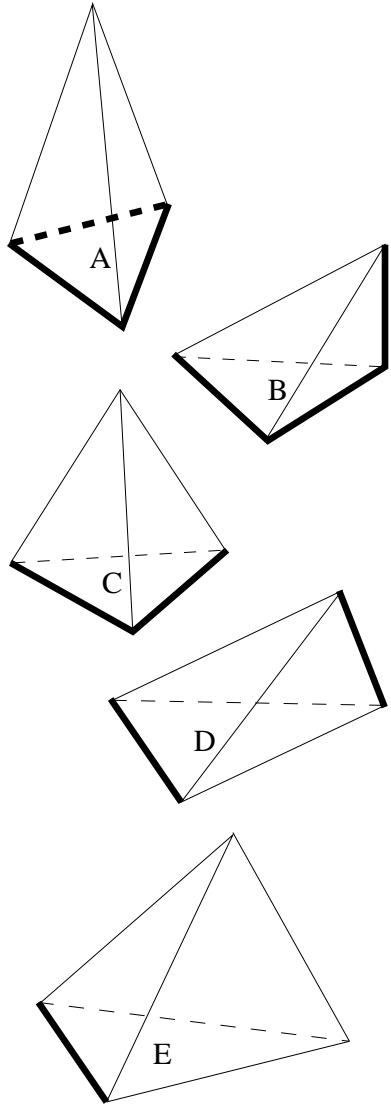}}
\caption{The five distinct two-parameter anisotropic tetrahedra. Edges
with length $l_{X}$ are shown in bold, the remaining edges are of length
$l_{Y}$.}
\label{fig:4}
\end{figure}

\begin{figure}[htbp]
\epsfxsize=3.10in
\centerline{\epsfbox[-10 175 580 835]{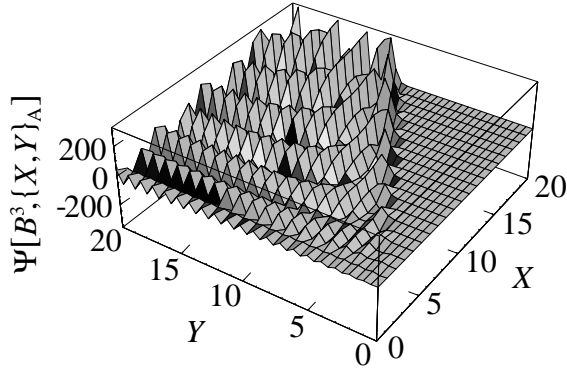}}
\caption{The type A wavefunction $\Psi[B^3,\{ X,Y\}_{\rm{A}}]$ as a function of
$X$ and $Y$ for ${0}\le{X}\le{20}$, ${0}\le{Y}\le{20}$. Observe that the
wavefunction vanishes where ${{X}\over{Y}}>2$ corresponds to an
inadmissible configuration.}
\label{fig:5}
\end{figure}

\begin{figure}[htbp]
\epsfxsize=3.0in
\centerline{\epsfbox[10 100 600 860]{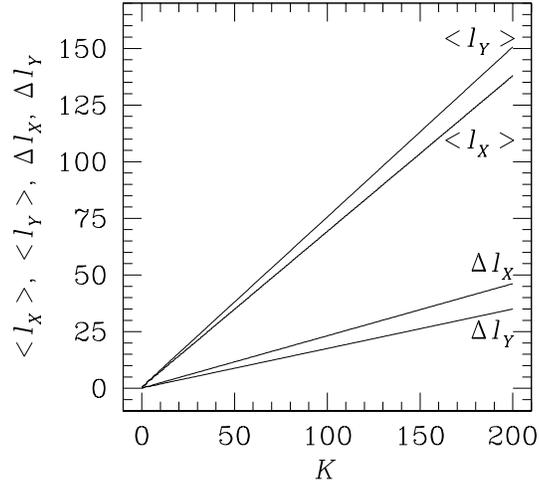}}
\caption{$<l_{X}>$, ${\Delta}l_{X}$, ${{<{l_{Y}}>}}$ and
${{{\Delta}{l_{Y}}}}$ for the type A tetrahedron for
${0}\le{K}\le{200}$.}
\label{fig:6}
\end{figure}

 
\begin{figure}[htbp]
\epsfxsize=6.0in
\centerline{\epsfbox[10 100 600 860]{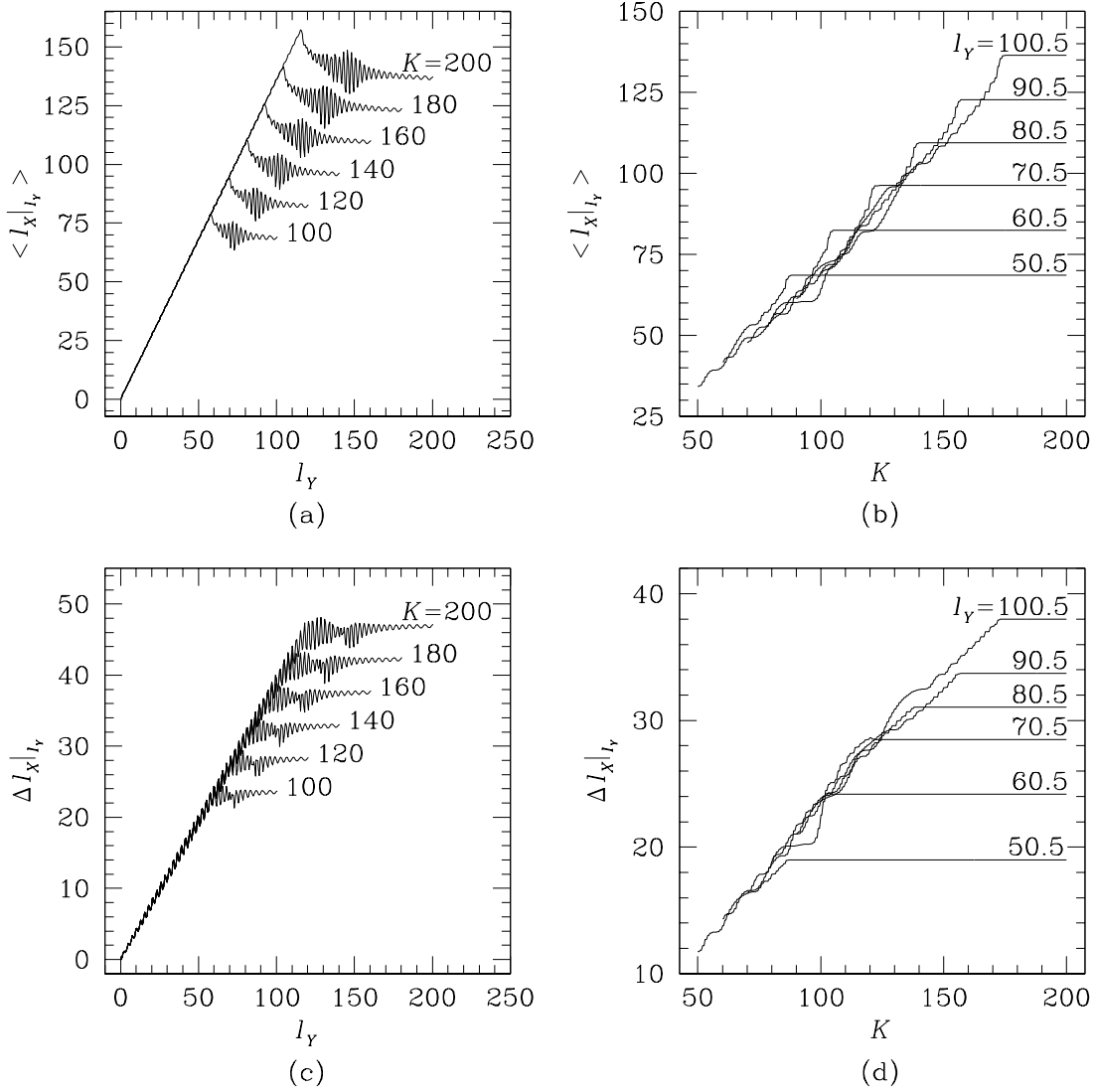}}
\caption{(a) $<l_{X}|_{l_{Y}}>$ and (c) $\Delta l_{X}|_{l_{Y}}$ for the
type A tetrahedron as functions of $l_{Y}$ for ${0.5}\le{l_{Y}}\le{(K+0.5)}$
for $K$ = 100, 120, 140, 160, 180 and 200; (b) $<l_{X}|_{l_{Y}}>$ and
(d) $\Delta l_{X}|_{l_{Y}}$ for the type A tetrahedron as functions of $K$
for ${(l_{Y}-0.5)}\le{K}\le{200}$ for $l_{Y}$ = 50.5, 60.5, 70.5, 80.5, 90.5
and 100.5.}
\label{fig:7}
\end{figure}

\begin{figure}[htbp]
\epsfxsize=6.0in
\centerline{\epsfbox[10 100 600 860]{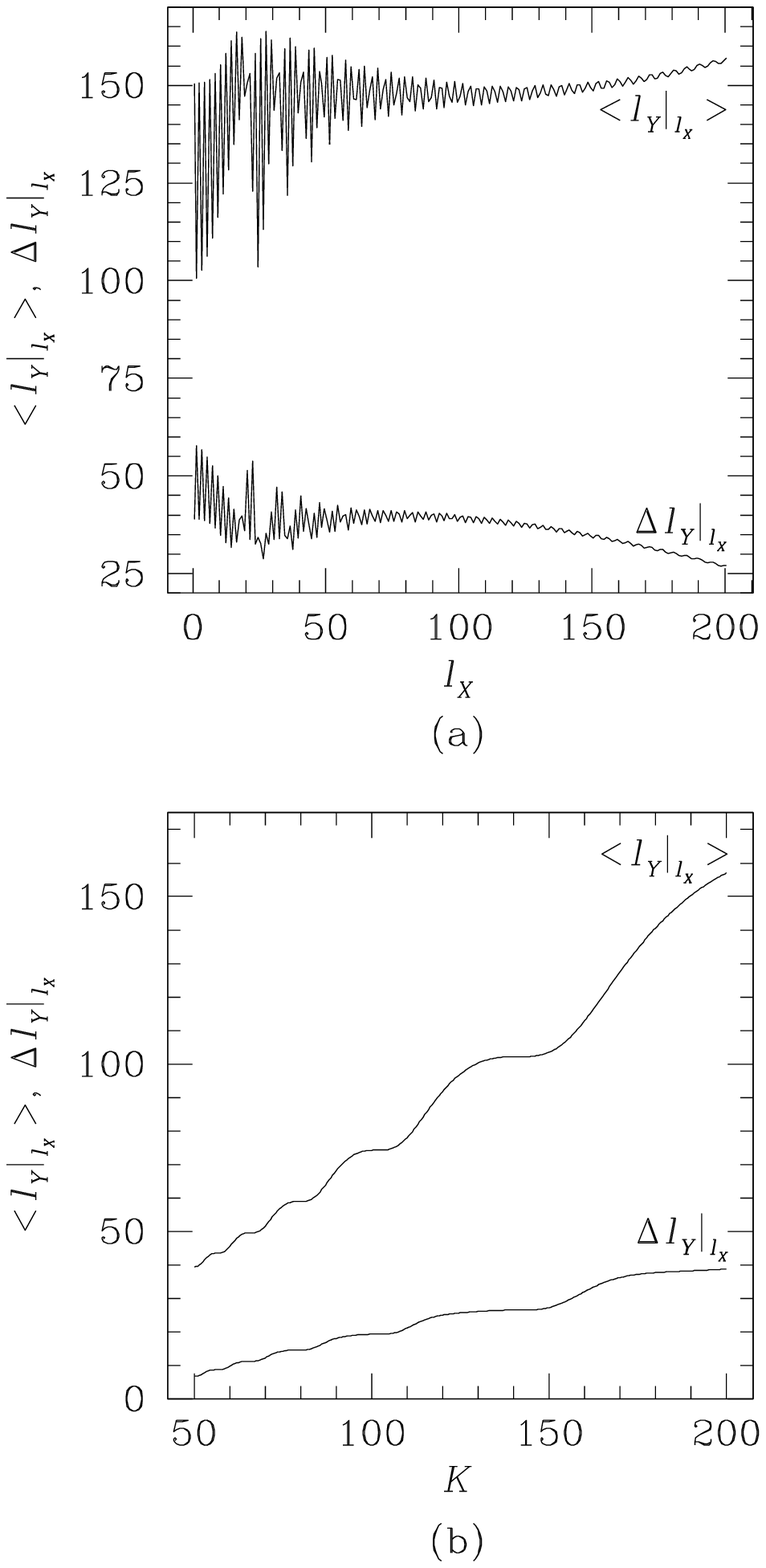}}
\caption{(a) $<l_{Y}|_{l_{X}}>$ as a function of $l_{X}$ for the type A
tetrahedron for ${0.5}\le{l_{X}}\le{200.5}$ and $K$ = 200; (b)
$<l_{Y}|_{l_{X}}>$ as a function of $K$ for the type A tetrahedron for
${50}\le{K}\le{200}$ and $l_{X}$ = 50.5. The expectation values and
uncertainties are clearly not linear.}
\label{fig:8}
\end{figure}

\begin{figure}[htbp]
\epsfxsize=6.0in
\centerline{\epsfbox[10 100 600 860]{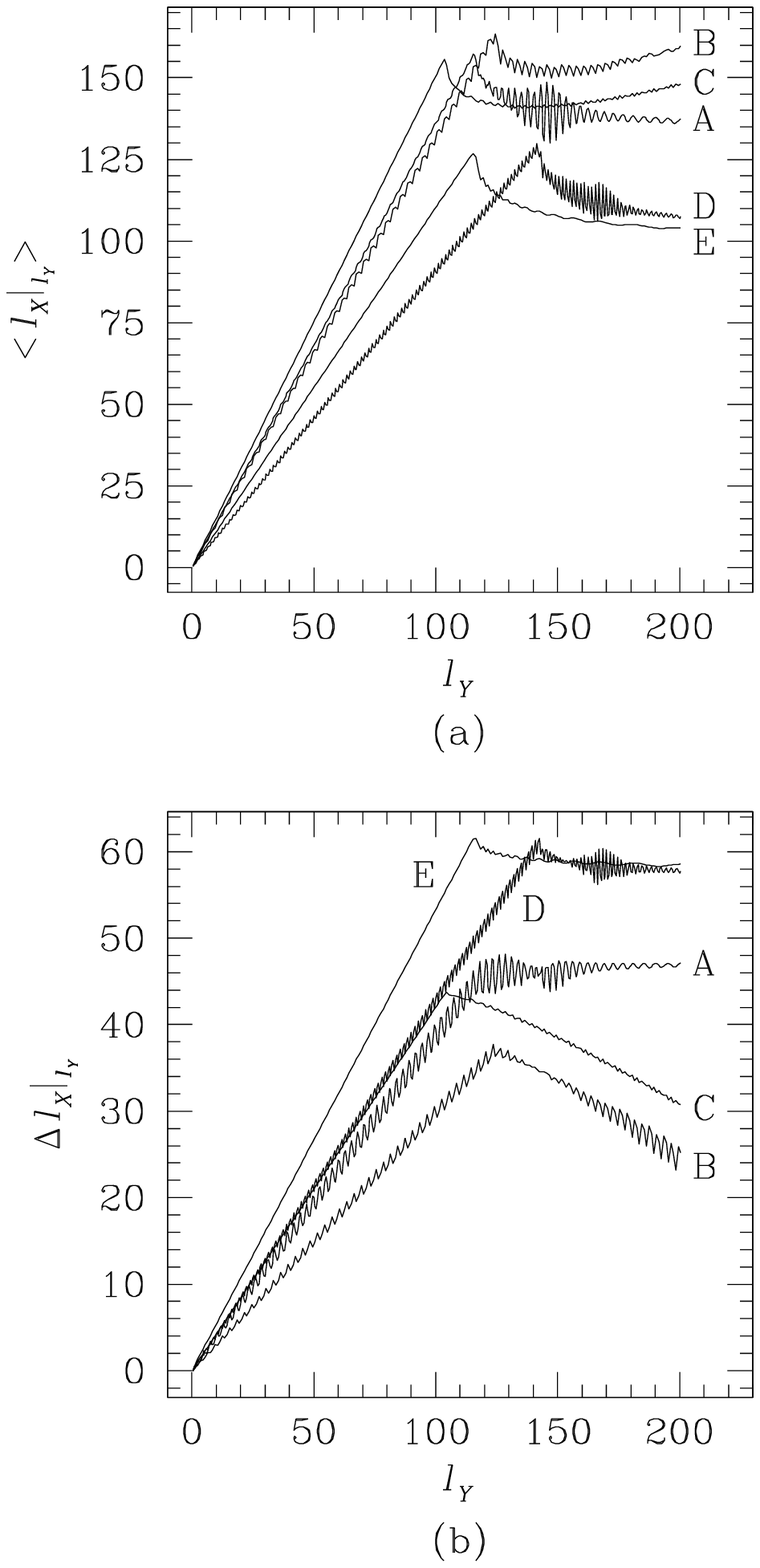}}
\caption{(a) $<l_{X}|_{l_{Y}}>$ and (b) $\Delta l_{X}|_{l_{Y}}$  as
functions of $l_{Y}$ for the five types of anisotropic tetrahedra for
${0.5}\le{l_{Y}}\le{200.5}$ and $K$ = 200.}
\label{fig:9}
\end{figure}

 
\begin{figure}[htbp]
\epsfxsize=6.0in
\centerline{\epsfbox[10 100 600 860]{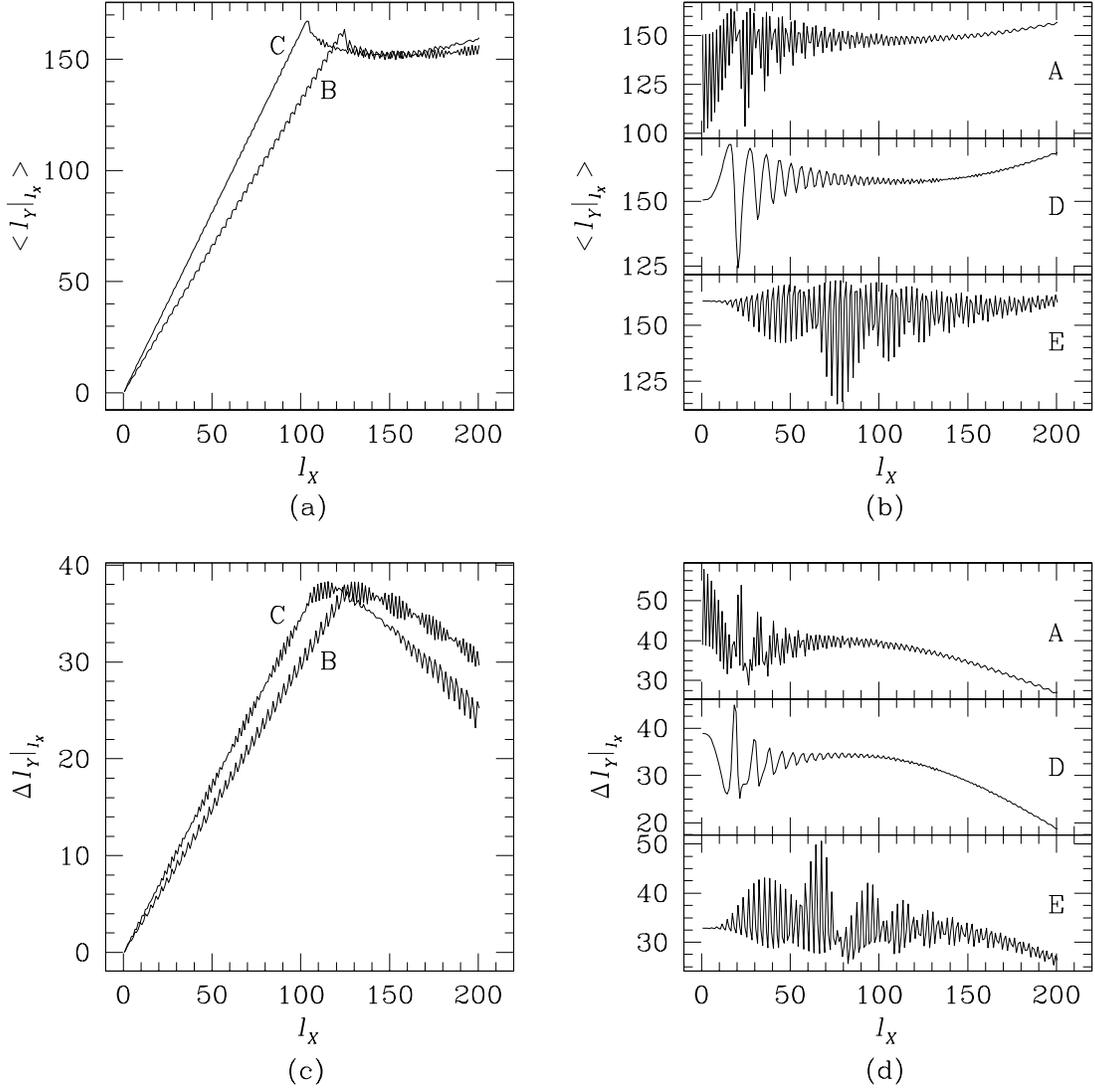}}
\caption{(a) $<l_{Y}|_{l_{X}}>$ and (c) $\Delta l_{Y}|_{l_{X}}$ as
functions of $l_{X}$ for the type B and C anisotropic tetrahedra for
${0.5}\le{l_{X}}\le{200.5}$ and $K$ = 200; (b) $<l_{Y}|_{l_{X}}>$ and (d)
$\Delta l_{Y}|_{l_{X}}$ as functions of $l_{X}$ for the type A, D and E
anisotropic tetrahedra for ${0.5}\le{l_{X}}\le{200.5}$ and $K$ = 200 are
clearly not linear.}
\label{fig:10}
\end{figure}

\end{document}